\documentclass{appolb}
\usepackage{epsfig}

\begin{document}
\pagestyle{plain}
\newcount\eLiNe\eLiNe=\inputlineno\advance\eLiNe by -1
\title{Emerging communities in networks - a flow of ties
\thanks{Presented on the 27-th Marian Smoluchowski Symposium on Statistical Physics, Zakopane, Poland, September 22-26, 2014. Correspondence to: {\tt kulakowski@fis.agh.edu.pl}}
}
\author{ Przemys{\l}aw Gawro\'nski, Ma{\l}gorzata J. Krawczyk and Krzysztof Ku{\l}akowski
\address{Faculty of Physics and Applied Computer Science, AGH University of Science and Technology, al. Mickiewicza 30, 30-059 Krak\'ow, Poland
}}

\maketitle
\begin{abstract}
Algorithms for search of communities in networks usually consist discrete variations of links. Here we discuss a flow method, driven by a set of differential equations. 
Two examples are demonstrated in detail. First is a partition of a signed graph into two parts, where the proposed equations are interpreted in terms of removal of a cognitive 
dissonance by agents placed in the network nodes. There, the signs and values of links refer to positive or negative interpersonal relationships of different strength. Second is an 
application of a method akin to the previous one, dedicated to communities identification, to the Sierpi\'nski triangle of finite size. During the time evolution, the related 
graphs are weighted; yet at the end the discrete character of links is restored. In the case of the Sierpi\'nski triangle, the method is supplemented by adding a small noise to 
the initial connectivity matrix. By breaking the symmetry of the network, this allows to a successful handling of overlapping nodes.
\end{abstract}
  
\section{Introduction}

Reliable and fast methods of identification of communities in networks are of interest for numerous applications in potentially all branches of knowledge, from biology 
to computer sciences \cite{boc,sfo}. For large networks, the condition of speed is crucial; on the other hand, to check all possible partitions is a NP-complete problem \cite{sfo}. 
Further, the definition of communities as 'groups of nodes within which connections are dense, and between which connections are 
sparser' \cite{boc}, although intuitively plausible, remains fuzzy. A quantitative method to distinguish between possible partitions of a given network is to calculate 
the so-called modularity $Q$ for all considered solutions \cite{memg} ; the one with the largest modularity is the proper one. Yet, to check that a partition is the proper one, we have 
to find it at first. \\

The purpose of this work is to report an idea that a social system, driven by a designed time evolution, is going to find the optimal partition by one's own. We mean that the algorithm of solving the problem is equivalent to modelling the actual dynamics of the considered system. The method is 
to solve numerically a set of $N(N-1)/2$ differential equations, where $N$ is the number of nodes of the network. Each equation is devoted to one element of the connectivity matrix. 
Solving differential equations numerically is computationally costly, hence rarely used, with \cite{gud} as an exception. The advantage is that the method leads deterministically 
to the sought after solution, i.e. to the partition most close to the initial state. \\

Below, two variants of the proposed equations are reported. In the first variant, the matrix elements $x_{ij}$ are related to social contacts between $N$ persons. The intensity 
and character (friendly or hostile) of these contacts are given as absolute values and signs of $x_{ij}$. The obtained partition of the group is interpreted as the solution of the 
Heider balance problem. In a nutshell, the idea can be reported as follows. In a seminal paper \cite{hei}, Fritz Heider has established the balanced and unbalanced configurations 
or mutual relationships of a triad of persons: in the former, the product of three related relationships is positive, and in the latter it is negative. This work was generalized to a network by Frank Harary \cite{har}, who has proved that the balanced state demands a clear division of the whole network into two groups, with friendly relationships within the groups 
and hostile relationships between all members of different groups.  This concept has been supplemented by a dynamic aspect by Leon Festinger \cite{lfs}, who indicated the ways people remove the cognitive dissonance, caused by unbalanced relationships. More recently, discrete algorithms of obtaining the balanced state have been constructed by \cite{red1,mar}. However, these prescriptions 
were plagued by jammed states, where some unbalanced triads were present. Later our method of differential equations \cite{kk1,aip,kk2} was also investigated by Steven Strogatz and 
coworkers \cite{str}, and no jammed states have been found there. In parallel, more and more historical events have been described in terms of the Heider balance \cite{moo,red2}, 
with the formation of coalitions of European states before WWI as a canonical example. \\

In the second variant, a more general algorithm is described, which is appropriate to any number of communities - not just two. On the contrary to the previous one, all links are positive. An additional parameter $\beta$ is introduced; its purpose is to separate relevant links from irrelevant ones. In a social system, the time evolution of links - relations between individuals - is equivalent to an improvement the relation in a dyad if its both members have good relations with other persons, and to a deterioration of the relation in the dyad if the relations of both members with others are bad. The method is to be applied with different values of $\beta$, and the final result is selected as to obtain the largest value of the modularity index $Q$. Also, during the time evolution the system passes through a series of subsequent partitions; again the solution with largest $Q$ is to be selected. The method has been formulated and validated for dense and sparse graphs in \cite{mjk1,mjk2}. A comparison with the results of \cite{memg}
allowed to state that in many cases the new method prevails. Let us note that in both our variants, the time evolution of the network is entirely deterministic, with the actual data as an 
initial state.\\

In two subsequent sections, we describe examples of applications of the two variants of the equations. Both examples have been reported already in \cite{aip,mjk3}; here we give 
a more detailed description of all stages of calculation. The first variant of the method has been validated in \cite{aip}. New element here is to determine the range of parameters 
where the solution is stable. Section 3 is devoted to the second variant, where it is applied to a particularly difficult case - a symmetric fractal. New element here is the time dependence of the modularity in the presence of noise. In the last section we highlight the difficulty met in our case by the deterministic method, when the condition of symmetry is compared to an unstable fixed point of the system dynamics.

\section{Cognitive dissonance}

According to the concept of cognitive dissonance, persons involved in mutual social relationships tend to order them as to reach a consistent division of others into friends 
and enemies. This tendency, when expressed in terms of the symmetric connectivity matrix $x_{ij}$, means that the relationships tend to fulfil the condition $x_{ij}x_{jk}x_{ki}>0$ for 
each triad $i,j,k$ \cite{hei}. This means that either all relationships in a triad should be positive (friendly), or two of them should be negative (hostile) and one - positive. These two 
configurations are balanced, while the other two (one or three negative relationships) are unbalanced. \\

\begin{figure}  
 \includegraphics[angle=270,width=0.86\columnwidth]{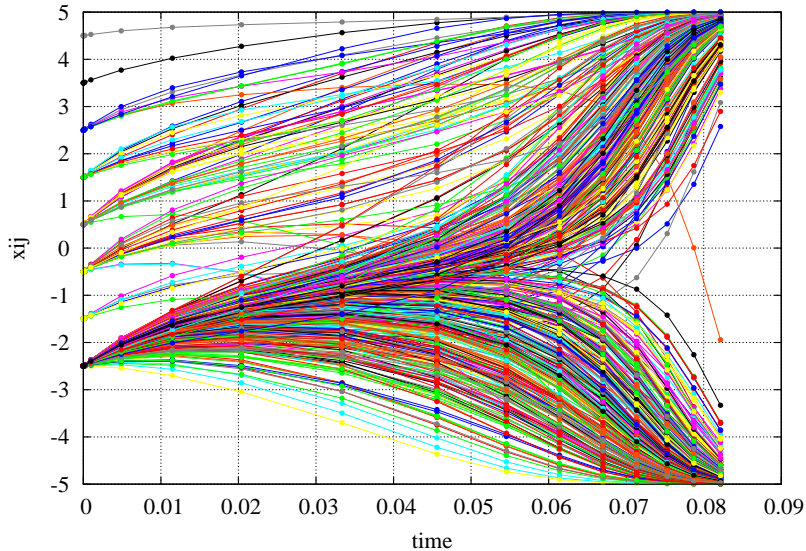} 
\caption{Time dependence of the relationships $x_{ij}$ driven by Eq. 2, with the elements of the matrix C (integers from 0 to 7) reduced by $\varepsilon=2.5$ as the initial values. 
The parameter $R=5.0$. When all relationships are balanced, the evolution is stopped.} 
\label{fig1}
\end{figure} 

\begin{figure}
 \includegraphics[angle=270,width=0.86\columnwidth]{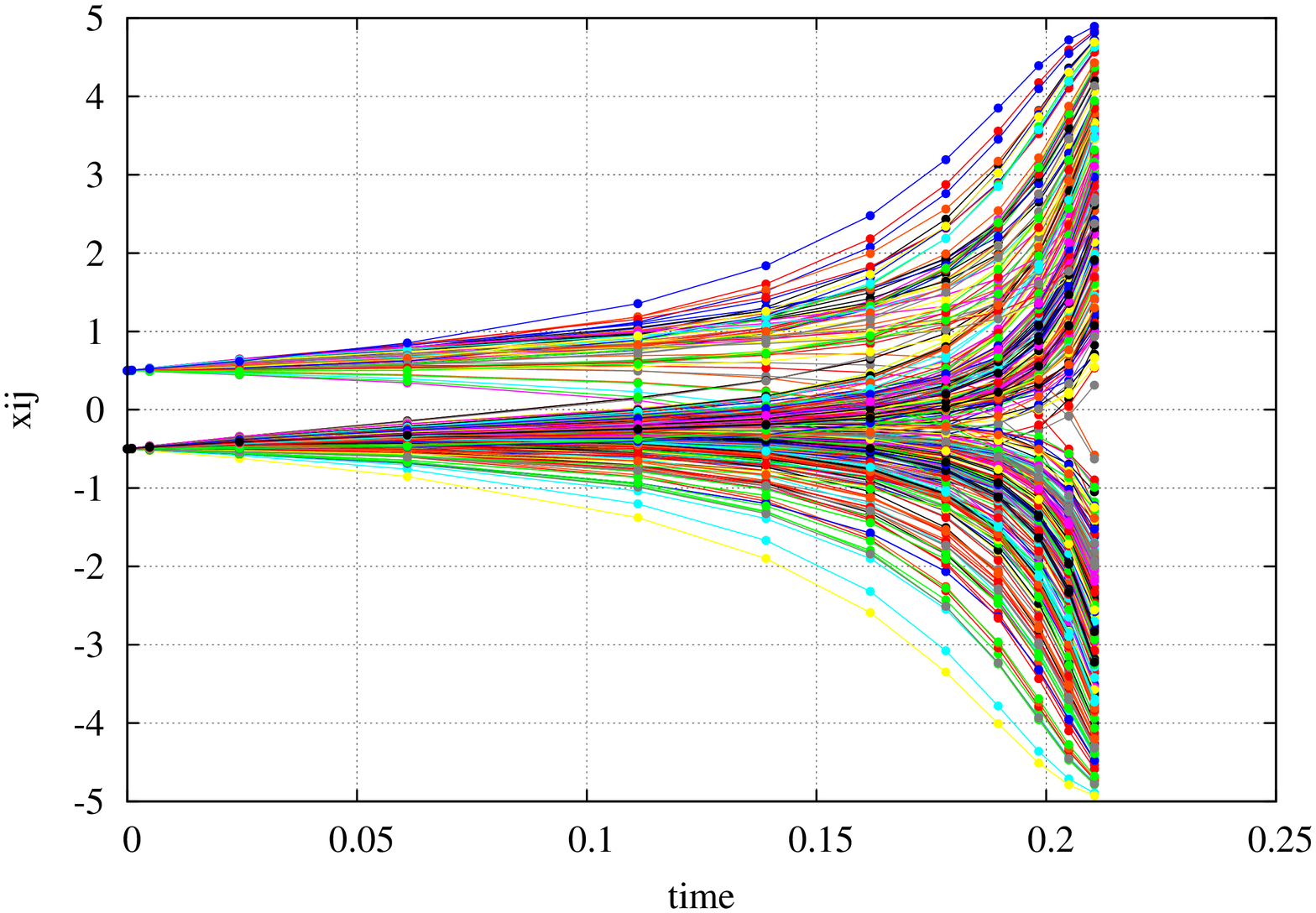} 
\caption{Time dependence of the relationships $x_{ij}$ driven by Eq. 2, with the elements of the matrix E (0 or 1) reduced by $\varepsilon=0.5$ as the initial values. The 
parameter $R=5.0$. When all relationships are balanced, the evolution is stopped.} 
\label{fig2}
\end{figure} 

If a network configuration is unbalanced, what should be the rule to remove the dissonance? The idea is taken from the set of four simple statements: {\it (a) We like someone
who likes someone that we like; (b) we like someone who dislikes someone we dislike; (c) we dislike someone who likes someone we dislike; and (d) we dislike someone who dislikes 
someone we like.} These statements are taken from the summary of the results of a careful laboratory experiment by Elliot Aronson \cite{aro}. The appropriate dynamics is that if 
A likes B and B likes D, then the relationship between A and D should go better. If, on the contrary, A likes B and B dislikes D, the relationship between A and D gets worse. 
Perhaps the simplest rule of this kind is

\begin{equation}
\frac{dx_{ij}}{dt}=\sum_{k\ne i,j} x_{ik}x_{kj}
\end{equation}
The flaw of this equation is that some matrix elements go to infinity in a finite time. This is seen for example for a triad, if all relationships $x,y,z$ are positive and initially 
equal; the solution is that $x(t)=x(0)/(1-x(0)t)$, and so is $y(t)$ and $z(t)$. A peculiar case is when all relationships are equal and negative; then all links increase and all should 
become friends. Yet, because of the symmetry, all matrix elements decrease equally; in the effect, the relationships become just more and more neutral (i.e. close to zero) 
when time goes to infinity. Both these symmetrical, then non-generic, cases can be easily generalized to larger networks; just the time should be multipled by the factor $(N-2)$, 
where $N$ is the number of nodes. Anyway, to evade the question how to interpret an infinite relationship, we limit the solutions $x_{ij}$ to the range $(R,-R)$, where $R>0$ is a 
parameter. This is done by the correction of the equations (1) as follows

\begin{equation}
\frac{dx_{ij}}{dt}=F(x_{ij})\sum_{k\ne i,j} x_{ik}x_{kj}
\end{equation}
where $F(x)=1-(x/R)^2$. Actually, the analytical results \cite{str} are obtained for the simplified version of equations, as in Eq. (1). In the non-generic symmetric case, when 
all relationships are initially of the same value, the marginally stable fixed point $x^*=0$ of Eq. (1) is supplemented by two new fixed points for Eq. (2): unstable $x^*=-R$ and 
stable $x^*=+R$. In agreement with this, for $x^*=-R$ all triads are unbalanced, while for $x^*=+R$ the whole network is balanced and all relationships are friendly. \\

Both for Eqn. (1) and (2), once the network is balanced, it remains balanced forever; this is so, because in the balanced state $sgn(x_{ij})=sgn(dx_{ij}/dt)$ for all links $ij$. This 
can be shown easily as follows: in each balanced triad, the product of three links is positive. Therefore, for a negative link, the product of the remaining two is negative; also, 
for a positive link, the product of the remaining two is also positive. These products contribute to the time derivative of the link, {\it q.e.d.} Below in this section we refer 
to numerical results obtained with Eq. (2).\\
 
The case to be reported here is the set of data on relationships between 34 members of an unspecified karate club, collected by Wayne Zachary in 70's \cite{zach}. Shortly after this 
search was performed, a conflict appeared between the club members, and the group was divided into two. These data, available in Internet \cite{vla}, are of common interest as a 
playground for all authors of algorithms designed to identify communities in networks. In particular, Mark Newman applied two own algorithms there \cite{memg,mej}. In the former
case, one member was assigned differently than in the reality; in the latter case, the obtained results exactly match the actual partition of the club members.\\

Actually, the data of Zachary are presented in \cite{vla} in two forms. In a more detailed version ('matrix C' in \cite{zach}), the graph is weighted: the values of the matrix 
elements - integers from 0 to 7 - reflect the relative strength/weakness of the relationships in the club. In a reduced version ('matrix E' in \cite{zach}), all non-zero values 
are substituted by 1's. We have applied Eq. 2 to both these forms. However, to get some links (relationships) negative, we reduced all matrix elements by the 
same value $\varepsilon$. For the matrix C, $\varepsilon=2.5$, while for the matrix E, $\varepsilon=0.5$. Both matrices are taken as the initial values of $x_{ij}$ for $34\times33/2=561$ 
differential equations. In both cases, the parameter  $R=5$. The obtained time dependences of $x_{ij}$ are shown in Fig. 1 (relationships weighted from 0 to 7, decreased by 
$\varepsilon=2.5$) and Fig. 2 (relationships reduced to 0 or 1, decreased by $\varepsilon=0.5$).\\

The result is that Eq. 2 reproduces the actual partition of the members, except the member No 9. (We note that in \cite{aip,kk3}, we erroneously wrote that our results exactly matched 
the real partition.) More surprisingly, our results for the matrices C and E are the same. This latter result is rather counterintuitive, because links equal to 1 or 2, classified as 
hostile in the matrix C, are converted to 1 in the matrix E, and then classified as friendly. Yet this means, that the solution is quite stable. \\

\section{Noisy fractals}

By definition, the Heider balance can be attained when either all relationships are friendly, or the network is divided into two parts. More generally, we need an equation which 
can lead to a division into any number of communities. The proposition \cite{mjk1,mjk2} is

\begin{equation}
\frac{dx_{ij}}{dt}=G(x_{ij})\sum_{k\ne i,j} (x_{ik}x_{kj}-\beta)
\end{equation}
where $G(x)=\Theta(x)\Theta(1-x)$ plays the same role as $F(x)$ in Eq. (2), and $\beta$ is a parameter. Here, all links $x_{ij}$ are positive; either they are weighted because a problem
formulation, or they were originally zero or one, and belong to the intermediate range $(0,1)$ only during the time evolution. The role of the parameter $\beta$ is to separate meaningful values of the 
product $x_{ik}x_{kj}$ from negligible ones. If this product is larger than $\beta$, the pair of links $(i,k)$ and $k,j$ contributes to an increase of the link $(i,j)$;
if the product is smaller than $\beta$, this pair contributes to a reduction of $x_{ij}$. The value of the parameter has to be found by trials; the criterion is to get the modularity 
$Q$ as large as possible. As the result of the time evolution, the modeled network is divided into more and more communities. The division which gives the largest $Q$ is accepted as 
final. \\

Here we are going to concentrate on an application of Eq. (3) to identification of communities in a finite Sierpi\'nski triangle. Namely, the network is formed from the 
nodes of the triangle, shown in Fig. 3. In \cite{mjk3}, the search of this structure has been combined with the method of system compression \cite{mjk4,mjk5}, applicable for symmetric 
systems. The Sierpi\'nski triangle was a suitable example. In Fig. 4 we show an exemplary time dependence of the modularity. For better visibility, first 2880 time steps are not shown. \\

It is easy to indicate in Fig. 3 nodes which should belong simultaneously to two communities; yet, to be clear, we indicate these nodes by arrows. Any procedure designed to assign each node to one community must leave these nodes as communities of one node. Yet, this cannot be a criterion of overlapping, because more communities of one node are possible. Indeed, the application of Eq. (3) gives the structure as follows. There are three communities of three nodes (1,2,4), (5,7,10), (9,11,13) and six communities of one node (0), (3), (6), (8), (12), (14). How to distinguish the overlapping nodes?\\

The idea is to add some noise to the elements of the connectivity matrix of the investigated structure. With noise, the symmetry is broken and an overlapping node can be assigned
to one or another community. Yet, the symmetry is preserved in the sense that the probabilities of assigning the node to two equivalent communities should be the same. Having performed 
the calculations many times, we can evaluate these probabilities. This is the criterion of overlapping nodes \cite{mjk3}.\\

The noise is introduced by adding small random numbers $\zeta$ to the elements of the connectivity matrix of value zero, and by subtracting $\zeta$ from the elements of value one.
The numbers $\zeta$ are different for each matrix element. Here they are drawn from the range $(0,a)$ with uniform distribution; $a$ is the amplitude of noise. Both $a$ and $\beta$
are free parameters, and the criterion to find their values is the maximization of the modularity $Q$. In Fig. 5 we show how the modularity $Q$, averaged over $10^4$ trials, depends 
on these parameters for $N=15$. As can be deduced from the presented plots, $\beta=0.15$ and the noise amplitude $a$ of the order of $0.01$ are appropriate. We note that the 
maximal $Q$ exceeds its average value. To give an example, for the Sierpi\'nski triangle of $N=15$ nodes and the optimal values of parameters given above, the maximal $Q$ is about 0.25. Yet, we must admit that this value is enhanced by some particular configuration of noise. More important result is that in the configurations which give this maximal $Q$, the nodes which are overlapping for $a=0$ are reasonably included in the communities nearby. For the triangle from Fig. 4, a typical partition which gives large $Q$ is 
(0,1,2,3,4), (5,7,8,10,12), (6,9,11,13,14). For this partition and zero noise, the modularity $Q$=0.26.\\

\begin{figure}
\includegraphics[angle=0,width=0.86\columnwidth]{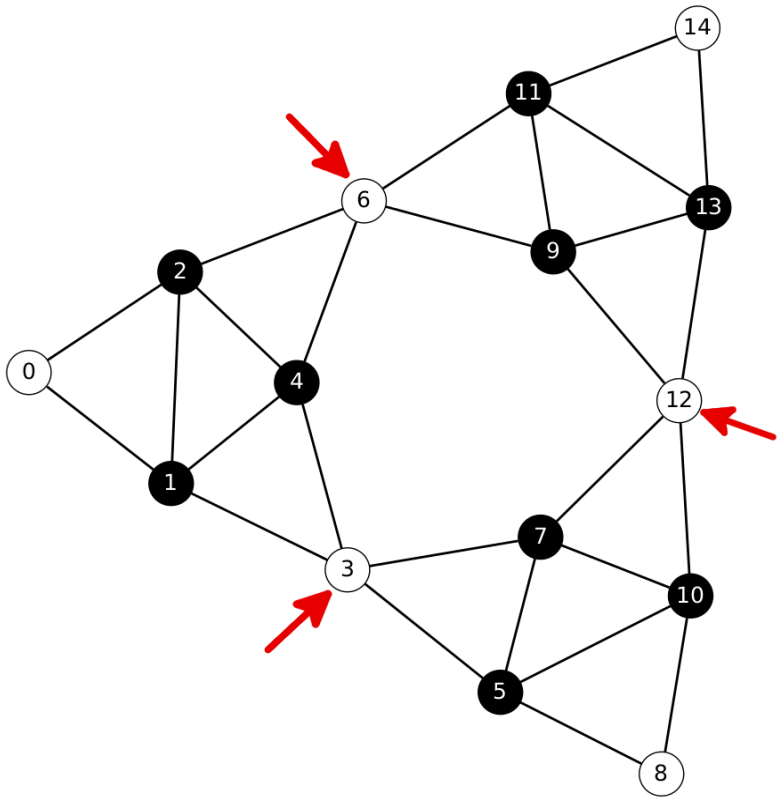} 
\caption{The Sierpi\'nski triangle of 15 nodes. Overlapping nodes are indicated by arrows.} 
\label{fig3}
\end{figure} 

\begin{figure}
\includegraphics[angle=270,width=0.86\columnwidth]{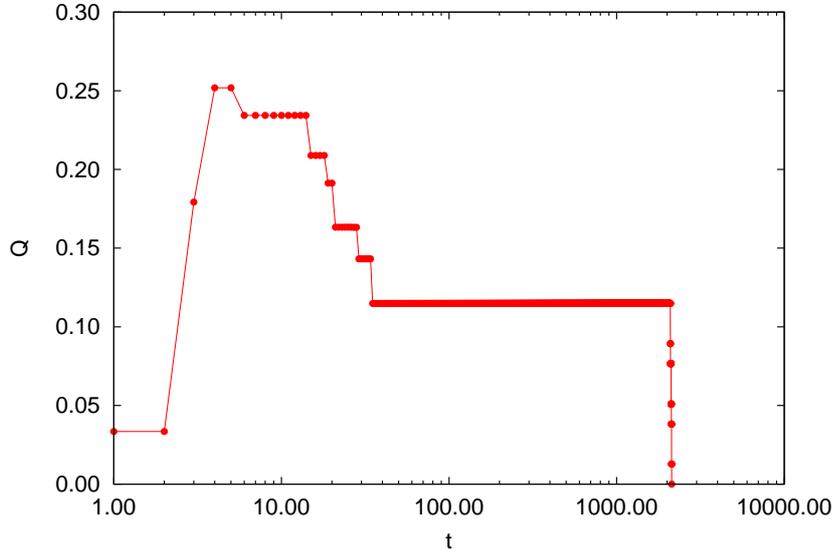} 
\caption{Time dependence of the modularity $Q$ - an example for the Sierpi\'nski triangle of 15 nodes.} 
\label{fig4}
\end{figure} 

\begin{figure}
\includegraphics[angle=270,width=0.86\columnwidth]{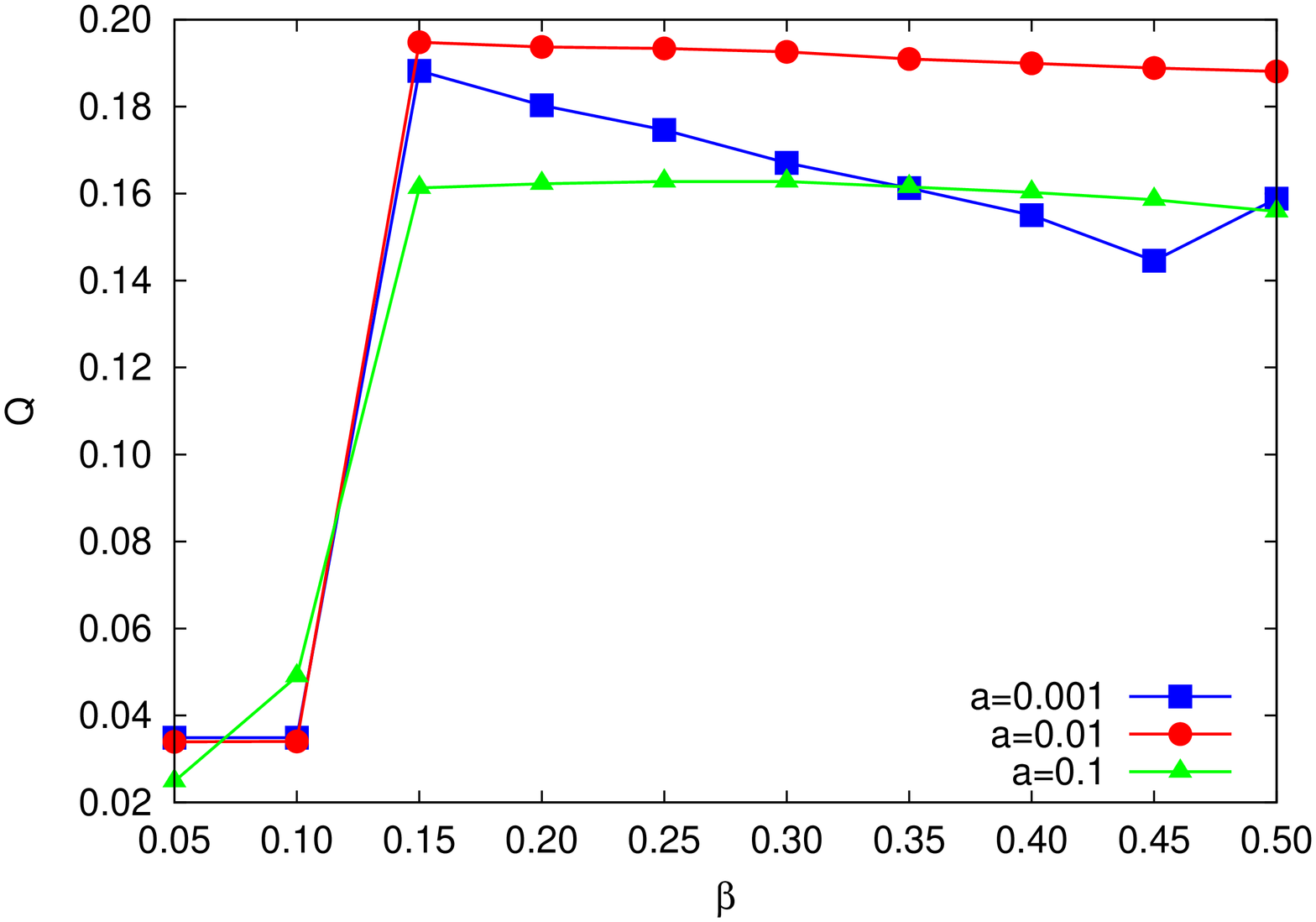} 
\caption{The modularity $Q$ as dependent on the parameter $\beta$ for the Sierpi\'nski triangle of 15 nodes, as in Fig. 3. Different curves are related to different amplitudes $a$ 
of the applied noise. The results are averaged over $10^4$ runs; for $10^3$ runs, the picture is practically the same.} 
\label{fig5}
\end{figure}

\section{Discussion}

To find a partition of a network is an inherently discrete task in the sense that all related variables are discrete. Both in the case of the Heider balance and in more general 
problem of communities, the task is to find a partition which is most close to the initial one. In the case of the Heider 
balance, the partition is such that positive links are only within the two parts of the network; these two parts are completely connected graphs with all links positive. All other links 
(between the parts) must be negative. In the more general case, where all links are non-negative, the condition is that all links between different parts of network are zero.
The idea to apply differential equations means that the transition from the initial to the final state is proceeding through a continuous variation of the links: they are represented
by integers at the beginning and end of the process, and by reals in the meantime. The advantage is that the process is deterministic: there is no randomness included. As found by 
\cite{str} for the case of the Heider balance, the dynamics driven by Eq. (2) always leads to the final state which obeys the demanded condition of balance, while an algorithm based 
on finite variations of discrete variables was shown to produce sometimes a jammed state. In terms of nonlinear dynamics, one could say that discrete algorithms based on Monte Carlo 
methods can produce solutions which are stable but unwanted, while differential equations are free from this flaw. The price we pay is the time of computation, much longer for differential equations  than for discrete algorithms, as in \cite{memg,mej}. Also, while in the variant of Heider balance the parameter $\varepsilon$ is almost irrelevant, in the 
variant of more communities the simulation should be repeated for different values of the parameter $\beta$, between 0 and 1. The accepted value of $\beta$ is the one which gives the largest modularity $Q$. Besides this, the algorithm is short and simple. \\

As we know from considerations of nonlinear dynamics, a system can be stuck in an unstable fixed point. We have seen a consequence of this in the case of the Heider balance, where 
the condition of symmetry (all matrix elements equal and negative) leads to a marginally stable fixed point where all links are zero. The condition of symmetry appears to be harmful 
also for the more general case. Namely, when we apply the method to a highly symmetric system, as the Sierpi\'nski triangle, the overlapping nodes are artificially isolated. This is so, 
because - by symmetry - they can be assigned neither to one, nor to another of the neighboring communities. Hence the role of noise, which breaks the symmetry and allows 
to obtain partitions with remarkably higher values of the modularity, than without noise.\\

\section*{Acknowledgements} One of the authors (K.K.) is grateful to the Organizers of the Smoluchowski Symposium for their kind hospitality. 
The research was partially supported by computing resources of ACC Cyfronet AGH and by the AGH UST project No. 10.10.220.01.

\end{document}